\documentclass[10pt,letterpaper]{article}
\usepackage{opex3,cite}
\usepackage{graphicx}
\usepackage{dcolumn}
\usepackage{bm,amstext}
\usepackage[colorlinks=true,breaklinks=false,hypertex]{hyperref}
\usepackage[normalem]{ulem}
\usepackage{amsmath,amssymb}

\newcommand{\kpp}{k^{(2)}}

\newcommand{\imm}{~{\rm mm^{-1}}}
\newcommand{\mic}{~\mu{\rm m}}
\begin{document}

\title{Generating mid-IR octave-spanning supercontinua and few-cycle pulses with solitons in phase-mismatched quadratic nonlinear crystals}

\author{Morten Bache, Hairun Guo and Binbin Zhou}
\address{$^1$DTU Fotonik, Department of Photonics Engineering, Technical University of Denmark, DK-2800 Kgs. Lyngby, Denmark}
\email{moba@fotonik.dtu.dk}

\begin{abstract}
We discuss a novel method for generating octave-spanning supercontinua and few-cycle pulses in the important mid-IR wavelength range. The technique relies on strongly phase-mismatched cascaded second-harmonic generation (SHG) in mid-IR nonlinear frequency conversion crystals. Importantly we here investigate the so-called noncritical SHG case, where no phase matching can be achieved but as a compensation the largest quadratic nonlinearities are exploited. A self-defocusing temporal soliton can be excited if the cascading nonlinearity is larger than the competing material self-focusing nonlinearity, and we define a suitable figure of merit to screen a wide range of mid-IR dielectric and semiconductor materials with large effective second-order nonlinearities $d_{\rm eff}$. The best candidates have simultaneously a large bandgap and a large $d_{\rm eff}$. We show selected realistic numerical examples using one of the promising crystals: in one case soliton pulse compression from 50 fs to 15 fs (1.5 cycles) at $3.0\mic$ is achieved, and at the same time a 3-cycle dispersive wave at $5.0\mic$ is formed that can be isolated using a long-pass filter. In another example we show that extremely broadband supercontinua can form spanning the near-IR to the end of the mid-IR (nearly 4 octaves).
\end{abstract}



\section{Introduction}
\label{sec:Introduction}
Ultrashort femtosecond mid-IR (wavelength range from $2.5-10\mic$) pulses is currently a hot topic since they can be used for probing \cite{bakker:2008} and even manipulating the ultra-fast dynamics of vibrational modes \cite{rini:2007}. These vibrational modes lie naturally in the mid-IR and by exciting them with mid-IR laser light one excites the fundamental vibrational modes instead of the overtones as with standard near-IR laser technology. In the ultrafast spectroscopy community there is therefore a request for energetic mid-IR pulses that are more broadband than what a commercial optical parametric amplifier (OPA) can deliver. This can be done in a specially designed OPA pumped by a near-IR laser \cite{Brida:2008,Ashihara:2009}, or through direct self-phase modulation (SPM) of the mid-IR beam exploiting the Kerr nonlinear refractive index in a bulk material \cite{Ashihara:2009-ol}; such pulses must be externally compressed in a dispersive medium to reach few-cycle duration. There is also a need for sources giving energetic and extremely broadband supercontinuum generation. Very broadband continua can be generated when pumping with mid-IR intense pulses either through SPM \cite{Corkum:1985} (i.e. before filamentation sets in) or by an increased peak power so filamentation sets in and a "white-light" continuum forms \cite{silva:2012}. Alternatively one may start with a femtosecond near-IR source and mix the frequency-converted harmonics in air and use four-wave mixing to achieve broadband mid-IR radiation \cite{Fuji:2007,Petersen:2010}, but with a very low yield. Currently energetic femtosecond mid-IR laser sources are developed using the optical parametric chirped pulse amplification technique \cite{Chalus:2010,Andriukaitis:2011} or regenerative amplification of mid-IR femtosecond lasers \cite{Moulton:2011}. Additional pulse compression to sub-two cycle duration of such sources is often desired to reach new frontiers in ultrafast and extreme nonlinear optics. In particular high-harmonic generation (HHG) has an extended harmonic cut-off when pumped with mid-IR front-ends due to a favorable $\lambda_0^2$ scaling \cite{corkum:1993}, that when combined with phase-matched HHG \cite{Popmintchev:2010} gives a decent yield even deep into the X-ray regime.  

Here we investigate whether ultra-fast cascaded optical nonlinearities can provide an alternative and more efficient route to facilitate these needs, or even be used as an additional component pumped by one of the recently developed mid-IR front ends to achieve pulse compression or spectral broadening as desired. We analyze a wide range of nonlinear frequency conversion crystals transparent in the mid-IR (both semiconductors and dielectric crystals) that can be used for ultrafast cascading. Numerical simulations are used to showcase the possibilities these crystals can offer for ultrafast mid-IR nonlinear optics. These crystals all carry very large $d_{\rm eff}$ values when pumping in the noncritical SHG configuration (where both pump and second harmonic have the same polarization). Until now they have not been investigated much for cascading because the phase mismatch is large and cannot be changed much, but as we will show they are very promising candidates for ultrafast mid-IR nonlinear optics.

\section{Background}

Cascaded SHG was suggested in the early days of nonlinear optics \cite{Ostrovskii:1967}, and it was experimentally shown to give rise to huge nonlinear phase shifts \cite{Thomas:1972,desalvo:1992} that can be controlled both in magnitude and sign through the phase mismatch parameter. In fact the effective cascading Kerr-like nonlinear index can be made several orders of magnitude larger than those of standard glass types \cite{sundheimer:1993}, and at the same time it gives access to a negative (self-defocusing) Kerr-like nonlinearity even far from any resonances, which is rather unique. In the process the frequency conversion from the fundamental wave (FW, $\omega_1$) to its second harmonic (SH, $\omega_2=2\omega_1$) is not phase matched $\Delta k=k_2-2k_1\neq 0$: after a coherence length $\pi/|\Delta k|$ only weak up-conversion occurs, after which back-conversion occurs after another coherence length. As this process is cyclically repeated (cascading) the FW effectively experiences a Kerr-like nonlinear refractive index change $\Delta n=n_{2,\rm casc} I$ proportional to its intensity $I$, and $ n_{2,\rm casc}\propto-d_{\rm eff}^2/\Delta k $, where $d_{\rm eff}$ is the effective quadratic nonlinearity. The total cubic nonlinear refractive index is then $ n_{2,\rm cubic}=n_{2,\rm casc}+n_{2,\rm Kerr}$, where  $ n_{2,\rm Kerr}$ is the self-focusing (positive) material Kerr nonlinearity. When $\Delta k>0$ and $|n_{2,\rm casc}|>n_{2,\rm Kerr}$, the total cubic nonlinearity becomes negative (self-defocusing), and self-focusing problems are avoided making the input pulse energy practically unlimited \cite{liu:1999}.

Moreover, both spectral broadening and temporal compression can occur in a single nonlinear material through solitons \cite{ashihara:2002}. Solitons are stable nonlinear waves that exist as a balance between nonlinearity and dispersion, which for a negative nonlinearity is achieved with normal dispersion. Thus, with cascading solitons can form in the near-IR where the majority of lasers operate and where most materials have normal dispersion.

Historically, \textit{critical} (type I) cascaded SHG has been used for energetic pulse compression \cite{liu:1999,ashihara:2002}. Despite not using the largest $d_{\rm eff}$-values, in critical SHG the birefringence is exploited to achieve $n_{2,\rm cubic}<0$ by angle-tuning the crystal close to phase-matching. However, when  $\Delta k<\Delta k_{\rm sr}=d_{12}^2/2\kpp_2$, where $d_{12}$ is the GVM parameter and $\kpp_2\equiv d^2k_2/d\omega^2|_{\omega=\omega_2}$ the SH GVD coefficient, this approach may result in a resonant (spectrally narrow) cascaded nonlinearity, impeding ultrafast applications \cite{bache:2007a} (see also discussion in \cite{zhou:2012}). Recently we showed that large cascaded nonlinearities can be obtained from a \textit{noncritical} (type 0) cascading without using quasi-phase matching (QPM) \cite{zhou:2012}.
The cascading nonlinearity is (a) large due to the huge $d_{\rm eff}$ of noncritical interaction, (b) always self-defocusing because $\Delta k>0$, (c) ultrafast because the cascading is non-resonant ($\Delta k>\Delta k_{\rm sr}$ turns out to always be fulfilled). We used this to show few-cycle soliton compression in the near-IR (1300 nm pump wavelength, sub-mJ pulse energy, 50 fs FWHM input pulse duration) in a just 1 mm long bulk LiNbO$_3$ crystal \cite{zhou:2012}. The input beam was collimated and loosely focused and thus propagated through the nonlinear crystal with minimal diffraction. After further propagation (using a 10 mm crystal instead of the 1 mm one) an octave-spanning supercontinuum was observed covering almost the entire near-IR range. In another publication we showed numerically that when a near-IR soliton is formed, few-cycle waves in the low-wavelength range of the mid-IR can be generated from the solitons \cite{bache:2011}.


Here we show how these techniques point towards really exciting applications in the mid-IR regime. The idea is that noncritical cascaded SHG may occur in bulk semiconductor or dielectric materials that are transparent in the mid-IR. Often these have huge quadratic nonlinearities, but tend to be overlooked for nonlinear frequency-conversion purposes because of the intrinsic lack of phase matching and because QPM techniques often cannot be applied or are immature technologies. The premise of our approach is the availability of an energetic, but multi-cycle, mid-IR pulse obtained e.g. by optical parametric amplification of a near-IR laser pulse, and then use cascaded SHG to invoke soliton compression towards few-cycle duration, and to create strong coherent spectral broadening. We stress that quasi-phase matching (QPM) is not used, and this is on purpose: it turns out that while QPM can increase $n_{2,\rm casc}$, it almost always leads to a resonant nonlinearity, which is a problem for ultrafast applications \cite{zhou:2012}.

\section{Soliton figure of merit: screening potential mid-IR crystals}

Looking for crystals supporting self-defocusing solitons requires $n_{2,\rm cubic}<0$, where the cascading nonlinearity is $n_{2,\rm casc}\simeq  -2\omega_1 d_{\rm eff}^2/c^2\varepsilon_0n_1^2n_2 \Delta k$. Using $d_{\rm eff}$ at a given wavelength, $\lambda_0$, and using Miller's scaling \cite{Ettoumi:2010} we get
\begin{align}\label{eq:n2-casc}
    n_{2,\rm casc}(\omega)
    &=-\frac{d_{\rm eff}^2(\omega)}{c\varepsilon_0n^2(\omega)n(2\omega)[n(2\omega)-n(\omega)]}\\ d_{\rm eff}(\omega)&=
    \frac{d_{\rm eff}(\omega_0)[n^2(\omega)-1]^2[n^2(2\omega)-1]} {[n^2(\omega_0)-1]^2[n^2(2\omega_0)-1]}
\end{align}
where $n(\omega)$ is the linear refractive index. We remind that in type 0 interaction the phase mismatch is always positive, and it only depends on wavelength (and somewhat also on temperature, which we here assume fixed). The Kerr nonlinearity is often not experimentally measured for the mid-IR crystals, so we choose to model the frequency dependence of $n_{2,\rm Kerr}$ (in SI units $\rm m^2/W$) using the two-band model (2BM) \cite{Sheik-Bahae:1998} as
$n_{2,\rm Kerr}(\omega)
=K'G_2(\hbar \omega/E_g)\sqrt{E_p}/n^2(\omega) E_g^4$,
where $K'$ is a material constant, fitted to $K'=7.3\times 10^{-9}~{\rm eV^{3.5}m^2/W}$ in wide-gap dielectrics \cite{desalvo:1996} and $K'=14.0\times 10^{-9}~{\rm eV^{3.5}m^2/W}$ in semiconductors \cite{Sheik-Bahae:1998}, $E_p=21$ eV is the Kane energy, which is constant for most materials, $E_g$ is the direct band gap energy, and the function $G_2(x)$ can be found in \cite[Table II]{Sheik-Bahae:1998}.

\begin{figure}[t]
  \begin{center}
    \includegraphics[width=10cm]{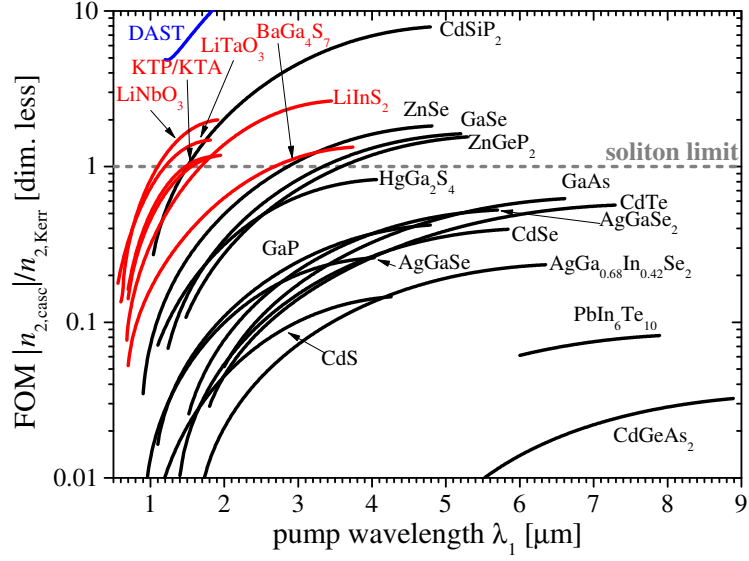}
  \end{center}
\vspace{-18pt}
\caption{\label{fig:cascading} Cascading figure-of-merit (FOM) $|n_{2,\rm casc}|/n_{2,\rm Kerr}$ vs. FW wavelength for cascaded noncritical type 0 SHG. Various dielectric (red) and semiconductor (black) materials are shown and the curves are only shown up to their $\lambda_{\rm ZD}$. A single organic crystal (DAST) is shown with blue. Sellmeier equations, $d_{\rm eff}$, and $E_g$ were taken from \cite{nikogosyan:2005,Kemlin2011,Das:2003,gayer:2008,shoji:1997,Petrov:2009,Avanesov2011,Santos-Ortiz:2011 ,Li:2000,Jazbinsek:2008,Jagannathan:2007,Lin2009}.
}
\end{figure}

The result of these calculations is shown in Fig. \ref{fig:cascading} as the figure-of-merit parameter $\text{FOM}=|n_{2,\rm casc}|/n_{2,\rm Kerr}$ vs. pump wavelength. This parameter must be larger than unity to observe self-defocusing solitons. We remind that the $K'$ parameter of the two-band model is found by fitting to experimental data, and is connected with some uncertainty, especially for dielectrics (materials plotted in red in the figure), but the convenience of representing the result with a ratio is that such an uncertainty will simply move the soliton limit either up or down. The well-known near-IR cascading materials LiNbO$_3$ and LiTaO$_3$ are seen to have an FOM up to 2, but at around $\lambda=2\mic$ the GVD changes sign and becomes anomalous (at which point the lines are interrupted; beyond $\lambda_{\rm ZD}$ we cannot excite self-defocusing solitons). The chalcogenide LiInS$_2$ and the semiconductors GaSe, CdSiP$_2$ and ZnGeP$_2$ are seen to have potential as well for $\text{FOM}> 1$, and this is because they simultaneously have large band gaps and large quadratic nonlinearities. (Note that the latter two are of the point group $\bar{4}2m$, giving an $ee\rightarrow e$ interaction of $d_{\rm eff}=3d_{36}\sin\theta \cos^2\theta$ that is maximum for $\theta=35^\circ$ and thus does not follow the traditional choice in type 0 interaction of polarizing along one of the optical axes.) CdGeAs$_2$ is an example of large quadratic nonlinearities and normal dispersion for long wavelengths, but also a very  small band gap, and therefore $n_{2,\rm Kerr}$ dominates due to the unfortunate $E_g^{-4}$ scaling. Another is GaAs, which was recently shown to have an overall self-focusing nonlinearity when pumped at $5.0\mic$ \cite{Ashihara:2009-ol}: cascading was found to be much smaller than the material Kerr nonlinearity, i.e. an FOM$\ll 1$, consistent with our results. Finally we also mention that organic materials can have huge $d_{\rm eff}$ values, e.g. over 1000 pm/V in DAST \cite{Jazbinsek:2008}, and can therefore score a very high FOM. However, it is not presently clear whether the 2BM can be used to predict the Kerr nonlinearity for these materials and therefore the high FOM for DAST should be considered uncertain.

\begin{figure}[t]
  \begin{center}
    \includegraphics[width=9cm]{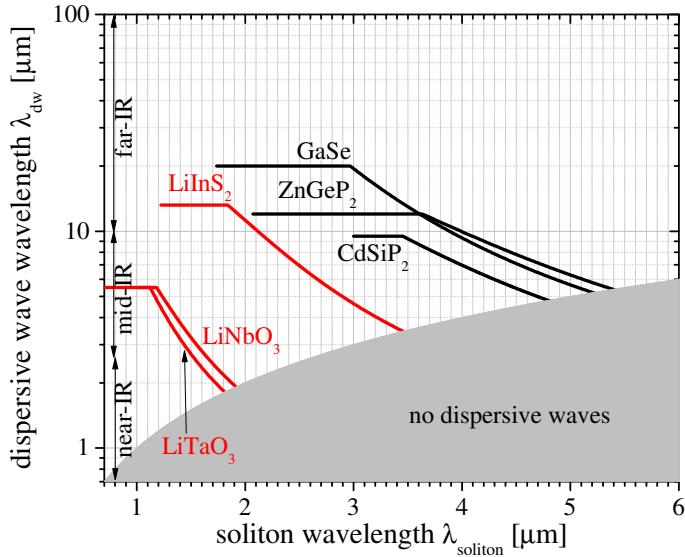}
  \end{center}
\vspace{-18pt}
\caption{\label{fig:dw-pm} The dispersive wave phase-matching point vs. wavelength of the self-defocusing soliton for selected crystals. In the gray area the dispersive waves cannot be excited because they would require a self-focusing soliton.}
\end{figure}

Let us briefly discuss how to excite a self-defocusing soliton once we have determined that the nonlinear crystal has FOM$>1$. When the FOM is above unity we have as mentioned before that $n_{2,\rm cubic}=n_{2,\rm casc}+n_{2,\rm Kerr}<0$. This is a necessary but not sufficient requirement to excite a soliton. The additional sufficient condition is that the \textit{effective }soliton order is unity or higher \cite{bache:2007}; it is defined as $N_{\rm eff}^2=|n_{2,\rm cubic}|I_{\rm in}L_D\omega_1/c$, where $L_D=T_0^2/|\kpp_1|$ is the dispersion length related to the FW GVD coefficient $\kpp_1$ and $T_0$ is the input pulse duration. Since the SHG interaction in the cascading limit ($\Delta kL\gg 2\pi$) essentially can be described by a nonlinear Schr{\"o}dinger equation (NLSE) governing the FW pulse dynamics with an effective cubic Kerr nonlinearity given by the coefficient $n_{2,\rm cubic}$, the soliton dynamics follow rather precisely the classical description for a fiber soliton (see e.g. \cite{agrawal:2007}), except the sign of the nonlinearity and GVD are opposite: For a unity effective soliton order the cascaded SHG supports a sech-shaped soliton with amplitude $\propto {\rm sech}(t/T_0)$. Exciting the soliton requires normal GVD ($\kpp_1>0$) as opposed to anomalous GVD in the classical soliton case. When increasing the soliton order the soliton will initially experience a self-compression stage (resulting in soliton compression of the input pulse \cite{ashihara:2002}), followed by temporal pulse splitting. In the ideal case the soliton dynamics for large soliton orders is cyclic and repeats itself over a soliton period $z=\pi/2L_D$. However, for femtosecond solitons higher-order dispersion, Raman effects and self-steepening-like terms will create soliton fission after the self-compression point, so that a train of solitons eventually will form. This is reminiscent of the classical route to supercontinuum generation (see e.g. \cite{dudley:2006}), except that the so-called optical Cherenkov waves induced by the soliton (also known as dispersive waves) are found in the anomalous dispersion regime on the red side of the spectrum due to the soliton being located in the normal dispersion regime \cite{bache:2008,bache:2010e}.

None of the crystals with $\rm FOM>1$ support self-defocusing solitons beyond $\lambda=5.5\mic$. However, because the soliton-induced optical Cherenkov radiation generates linear dispersive waves in the anomalous dispersion regime beyond the zero-dispersion wavelength $\lambda>\lambda_{\rm ZD}$ with a high conversion efficiency (up to 25\%) \cite{bache:2011a}, generating few-cycle pulses in the upper end of the mid-IR regime can conveniently be covered in this way. To get an idea about the wavelengths that can be excited, we show the dispersive-wave phase-matching curves
\cite[Eq. (2)]{bache:2011a} vs. soliton wavelength in Fig. \ref{fig:dw-pm} for the most promising materials.

\begin{figure}[t]
  \begin{center}
    \includegraphics[height=4cm]{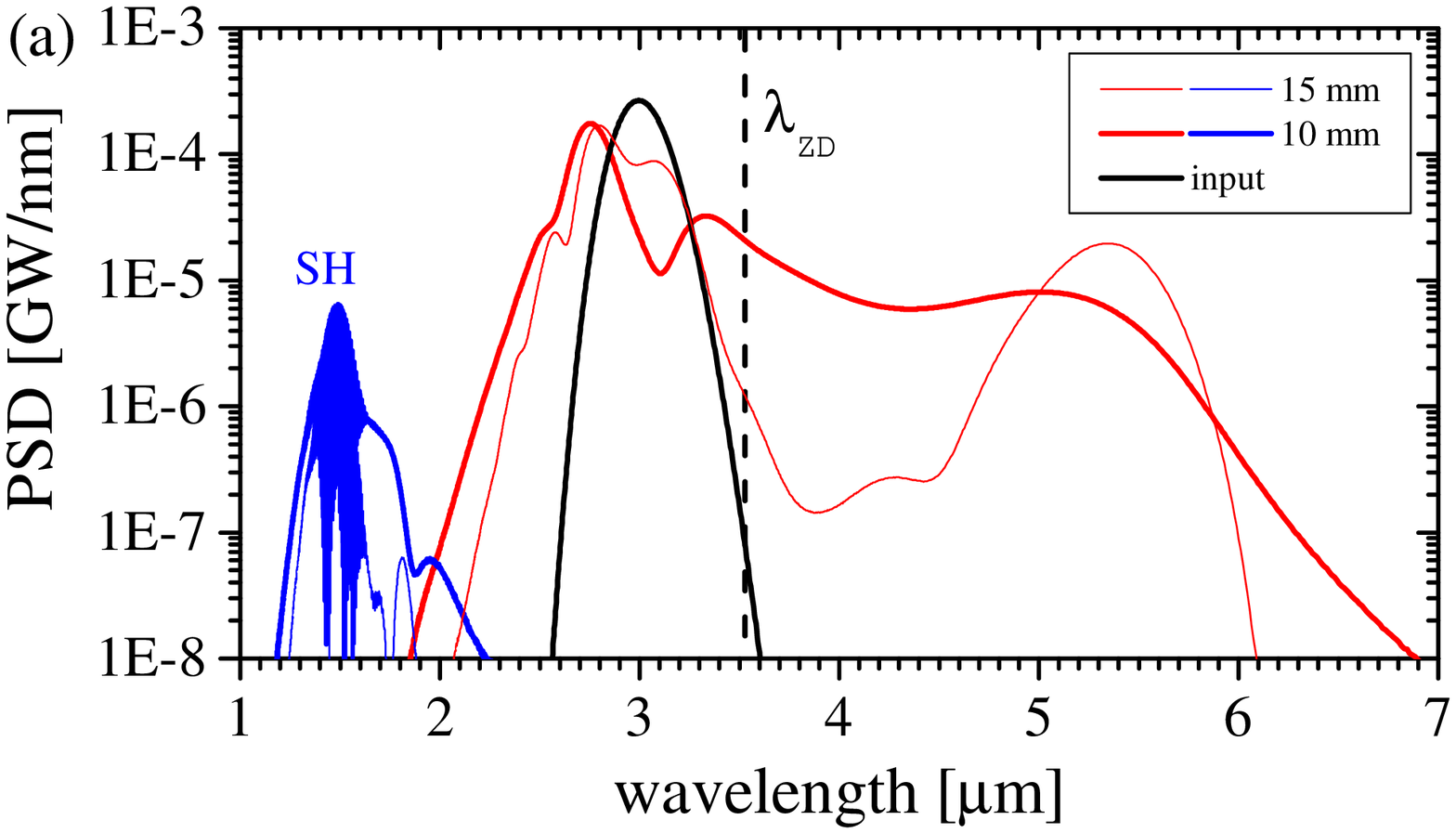}\\
    \includegraphics[height=3.5cm]{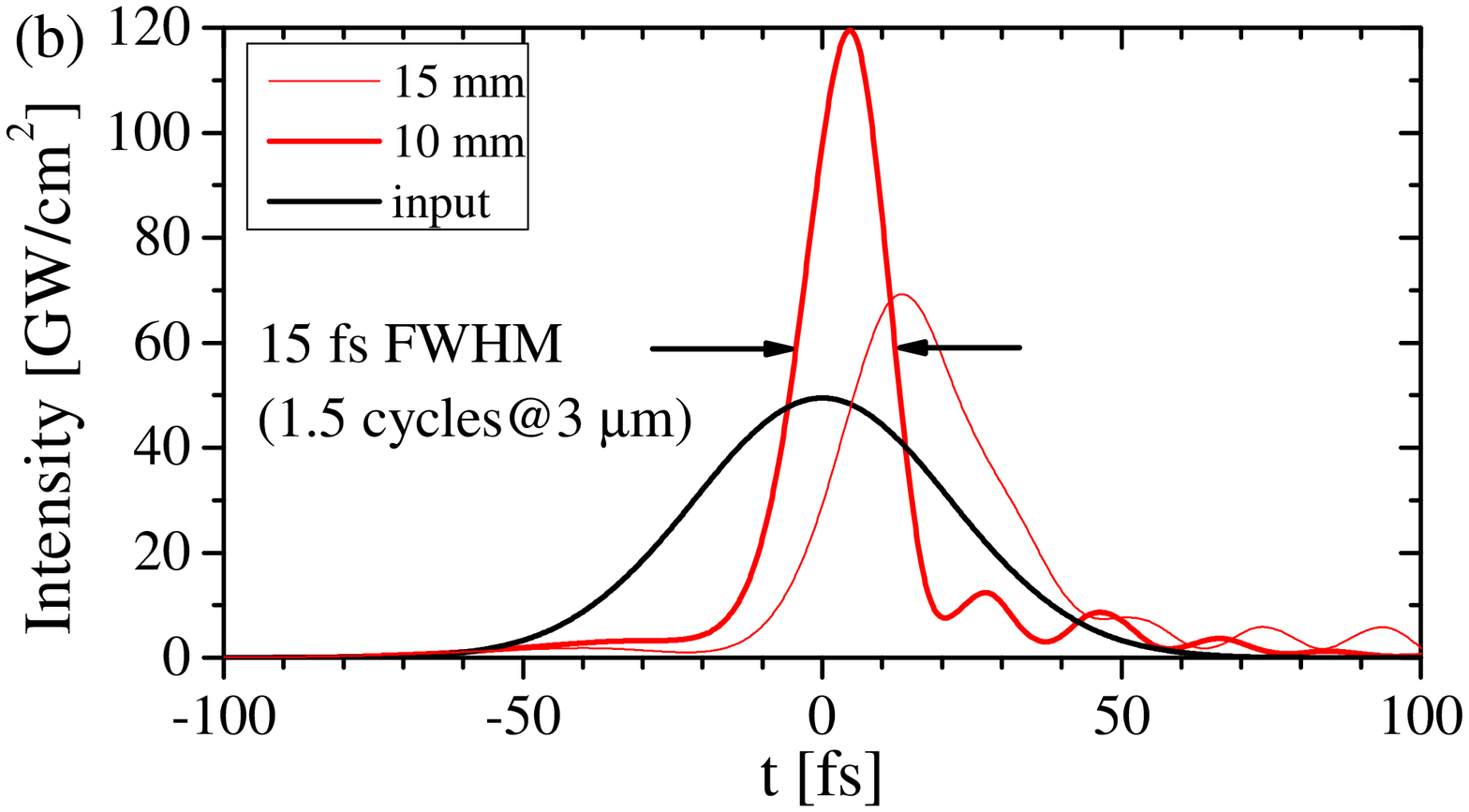}
    \includegraphics[height=3.5cm]{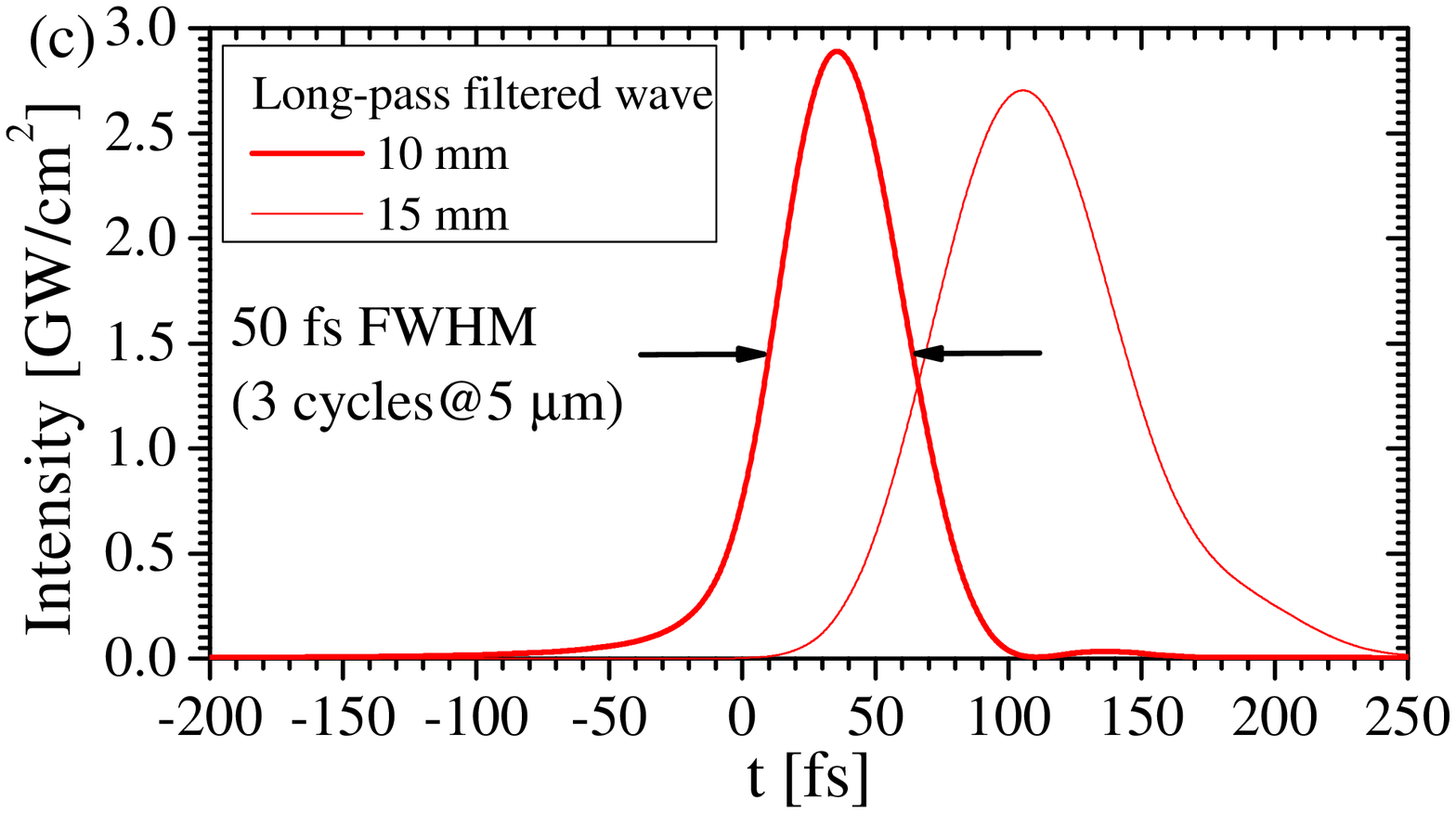}
  \end{center}
\vspace{-18pt}
\caption{\label{fig:sim-compression} Numerical simulation of soliton compression in a 15 mm long LiInS$_2$ bulk crystal. (a) Power spectral density at 0, 10 and 15 mm. (b) Temporal intensity at 0 and 10 mm. Input pulse: 50 fs FWHM and 50 GW/cm$^2$ centered at $\lambda_1=3.0\mic$. The simulation was done with a plane-wave SEWA model and the PSD was calculated assuming a $\simeq 0.5$ mm spot size, corresponding to an input pulse energy of around $6~\mu$J.  }
\end{figure}

\section{Numerical examples}

In the following we show numerical simulations using the LiInS$_2$ crystal as a typical example of the mid-IR performance. We chose this crystal as it is commercially available up to 15 mm length with a decent aperture, and it is broadband transmitting in the range from $0.5-9\mic$ ("0.5" level, \cite{Fossier:2004}). The simulations rely on the coupled slowly-evolving wave equations (SEWA) \cite{moses:2006,bache:2007} that are accurate to the single-cycle limit. We neglect diffraction for simplicity as the beam is assumed to be loosely focused and collimated along the crystal, and the overall nonlinearity is self-defocusing. The input fields are seeded with noise (one photon per mode \cite{dudley:2006,Frosz:2010}) where some assumption about the beam area must be taken. We include Raman effects, modeled as a single dominant Raman mode \cite{Fossier:2004} at 268$~\rm cm^{-1}$ using a Lorentzian having a temporal oscillation period of 20 fs and a relaxation time of 1200 fs (Lorentzian width of 8.5$~\rm cm^{-1}$). The crystal is a biaxial crystal in the $mm2$ point group. We consider a $Y$-cut ($\theta=90^\circ$, $\phi=0^\circ$) crystal resulting in a heavily phase-mismatched $ss\rightarrow s$ SHG interaction when the input light is polarized along the slow-index optical axis (hence the notation $s$-wave, with $Z=c$, refractive index given by $n_Z$). The interaction exploits the largest quadratic nonlinear tensor $d_{\rm eff}=d_{33}=16$ pm/V@$2.3\mic$ \cite{Fossier:2004}, and both FW and SH have the same polarization (noncritical type 0 SHG). The cascading strength around $3.0\mic$ is $n_{2,\rm casc}=-60\times 10^{-20}~\rm m^2/W$, while the electronic Kerr nonlinearity as calculated from the 2BM is roughly half that. This favorable FOM$\simeq 2$ comes in part from a large $d_{\rm eff}$ and a large bandgap $E_g=3.55$ eV \cite{Fossier:2004}. 

In Fig. \ref{fig:sim-compression} we show a numerical simulation using a typical pulse from a Ti:Sa pumped fs OPA, namely a 50 fs FWHM $\lambda=3.0\mic$. Taking a moderate intensity $I_{\rm in}=50~\rm GW/cm^2$ the simulation shows the result when the pulse is propagated in a 15 mm long LiInS$_2$ crystal, generating after 10 mm a sub-2-cycle compressed pulse by excitation of a self-defocusing soliton. This soliton radiates a broadband dispersive wave in the linear, anomalous dispersion regime, seen as the plateau located around $\lambda=5.0\mic$. Upon further propagation the dispersive wave grows in strength but becomes slightly more narrow in bandwidth. As shown by us recently \cite{bache:2011a}, we can employ a long-pass filter to separate the dispersive wave from the soliton around $\lambda=3\mic$. In fact, by simply removing all spectral content below $\lambda=4.5\mic$ we get the mid-IR pulses shown in (c): at 10 mm a 50 fs FWHM pulse is obtained, and considering the center wavelength of around $5\mic$ this corresponds to a 3-cycle pulse. After further propagation the pulse becomes slightly broader (around 80 fs FWHM or 4.5 cycles) because it disperses but it also becomes more energetic; at 10 mm around 6.5\% of the energy is located in the filtered wave while at 15 mm it is increased to 9\%. Such a conversion efficiency is quite high considering that a few-cycle mid-IR pulse is generated. 

\begin{figure}[t]
  \begin{center}
    \includegraphics[height=6cm]{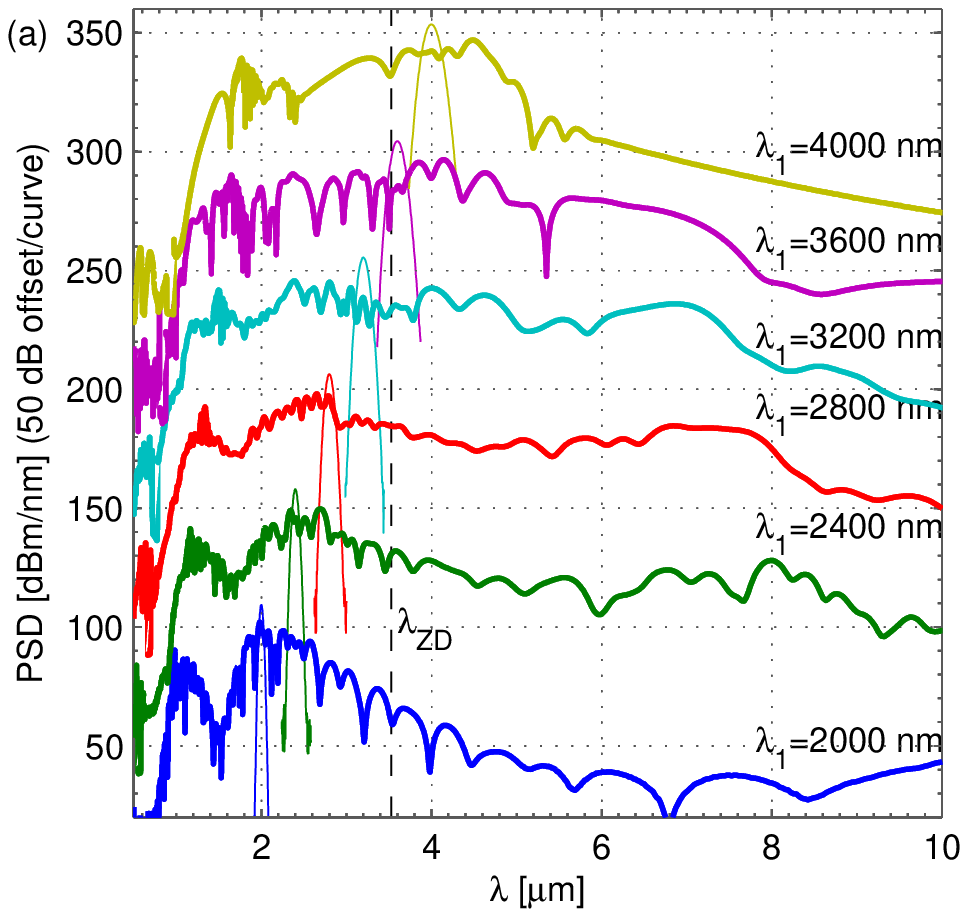}
    \includegraphics[height=6cm]{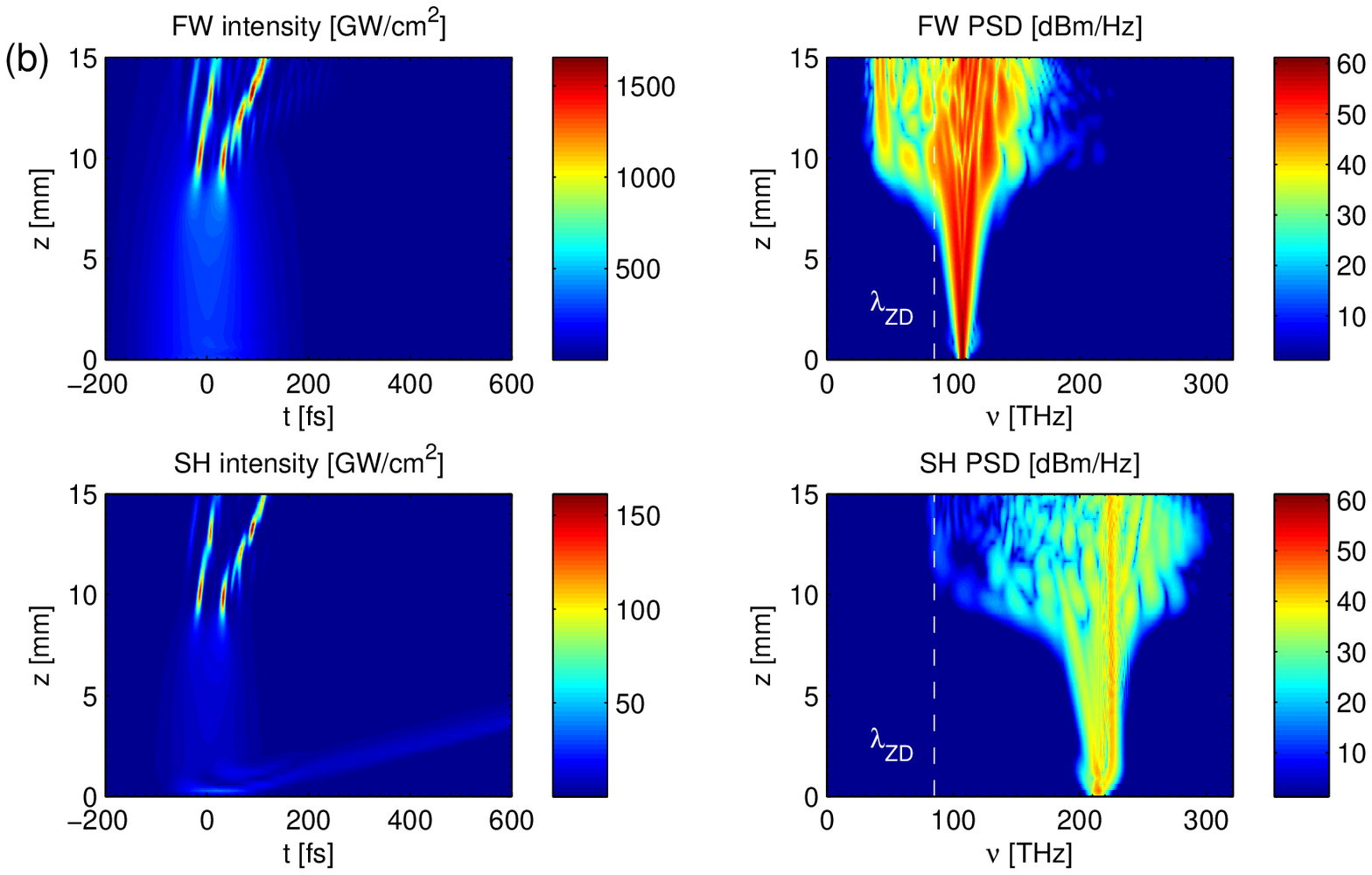}
  \end{center}
\vspace{-18pt}
\caption{\label{fig:sim-SCG} Numerical simulation of supercontinuum generation in a 15 mm long LiInS$_2$ bulk crystal. (a) Power spectral density of the output pulses (thick lines) for input wavelengths ranging from 2.0-4.0$\mic$ (thin lines). All input pulses have the same input intensity 300 GW/cm$^2$ and pulse duration 200 fs FWHM. The simulations were done with a plane-wave SEWA model and the PSD was calculated assuming a $\simeq 0.5$ mm spot size, corresponding to an input pulse energy of around $0.2~$mJ. The input pulses were seeded with noise (one photon per mode) and the PSDs are averaged over 50 realizations. Each PSD displays the combined FW and SH spectral content. (b) Single realization at $2.8\mic$ showing the pulse evolution along the crystal for the FW and SH intensities in time and the PSDs in frequency domain. }
\end{figure}

It is also interesting to pump with a longer pulse to investigate what happens when the soliton order becomes very high, giving substantial soliton fission after the initial self-compression stage and eventually to the generation of a broadband continuum. In the previous simulation the effective soliton order was around 2. By increasing the pulse duration to 200 fs FWHM and the intensity to 300 GW/cm$^2$ we achieve an effective soliton order that is around 10 times larger. The results of the simulations are summarized in Fig. \ref{fig:sim-SCG}(a), where we plot the power spectral density (PSD) vs. wavelength for pump pulses ranging from 2.0-4.0$\mic$, i.e. from well below and just above the zero dispersion wavelength. Note the PSD contains information about both the FW and SH, as they have the same polarization, which we discuss below. In all cases a very strong supercontinuum is generated, but in particular the three cases of $\lambda_1=2.4\mic$, $2.8\mic$ and $3.2\mic$ show a very broad supercontinuum, spanning basically the entire near-IR and lower mid-IR range from $1.0-8.0\mic$. The blue part of the spectrum is given by the contributions from the SH, see more in the discussion below. The reason why the broadest continua are found for these wavelengths is related to the fact that when approaching the zero dispersion wavelength the GVD drops and therefore very high soliton order is invoked for the chosen intensity ($\gg 10$). Conversely the GVD is much larger at $2.0\mic$ so in order to reach similar huge soliton orders would require input intensities on the order of 1.0 TW/cm$^2$, and it is not clear whether the crystal can tolerate such larger intensities. We also notice that in two of the cases the pump wavelength lies in the linear (anomalous) dispersion regime. The consequences is particularly clear for the 4000 nm case, where spectral ripples are observed, originating from optical wave breaking that occurs for high soliton orders when GVD and nonlinearity have the same sign \cite{Tomlinson:1985}.

In (b) we show the detailed temporal and spectral dynamics of a single realization at $\lambda_1=2.8\mic$. We see that the FW experiences a strong self-compression stage up until $z=10$ mm. Interestingly two solitons seem to form simultaneously here, somewhat reminiscent of a modulational instability that we believe stems from the Raman effects (cf. also \cite{zhou:2012}). After the compression point dispersive waves and soliton fission collide and interact in a complicated manner, and it is this kind of interaction that generates a broad continuum.

We also show the SH dynamics, and it basically comprises of two components: a GVM-induced pulse that quickly disappears as the SH group velocity is much lower than the FW. Secondly a very strong component is seen traveling with the FW group velocity and essentially contains the same dynamics as the FW: this is due to the slaving of the SH that occurs in the cascading limit \cite{bache:2007a}. The SH is often considered irrelevant in cascading due to the large phase mismatch that is inherently connected to the process. However, in the case considered here the phase mismatch is not that strong, $\Delta k=80\imm$ for $\lambda_1=2.8\mic$, and combining this with a high intensity quite a significant amount of energy lies in the SH. As is evident the spectral broadening is substantial since the SH essentially "copies" the FW spectrum due to the slaving. Since the SH in the noncritical interaction actually carries the same polarization as the pump, experimentally one would measure a spectrum spanning the entire range covered by the two spectra, and by inspecting the PSDs in Fig. \ref{fig:sim-SCG}(b) by eye alone this is a quite substantial range. We therefore decided in (a) to show the combined PSD of the FW and SH. It was calculated by reconstructing the physical (real) electrical fields from the SEWA envelopes \cite{brabec:1997}, adding them to get a total electrical field in time and then transform to frequency domain. Note that the approximations of the envelope model starts to break down once the harmonic envelopes overlap (substantially). We therefore checked that we could find essentially the same results using a more complete numerical model, which does not rely on modeling the envelope but rather the electrical field. It therefore models the FW and SH interaction as a single electrical field without decomposing the carrier wave into envelopes describing each harmonic. The technique is described in a recent paper \cite{guo:2013}.

Let us finally touch upon the issue of increasing the repetition rate and lower the pulse energy. This would require waveguiding of the pulses, and the techniques in mid-IR materials are still immature compared to the well-developed near-IR range. For the dielectric crystals ridge waveguides can be made on the surface with a blade-cutting technique or one can use femtosecond laser inscription by focusing a femtosecond beam below the surface and achieve an index contrast by material damage. The semiconductor materials are instead more compatible with clean-room technologies, so it is realistic to hope for good-quality waveguides in the near future. In a future publication we will go into detail about this topic.

\section{Conclusion}

Summarizing, we investigated mid-IR nonlinear crystals in the type 0 phase-matching configuration for ultrafast cascaded SHG. Several crystals showed potential for compressing multi-cycle, energetic mid-IR pulses towards few-cycle duration using self-defocusing solitons. While these solitons could only be found below the zero dispersion wavelength (ranging from $\lambda=2.0-5.5\mic$ for the crystals we investigated), the upper end of the mid-IR can also be covered because the solitons will generate ultrashort, broadband dispersive waves located beyond the zero dispersion wavelength. We used numerical simulations to highlight some specific examples: firstly a case where a $3.0\mic$ 50 fs pulse with a moderate intensity was compressed to 1.5 cycles. This occurred after 10 mm propagation in a bulk and heavily phase-mismatched mid-IR frequency conversion crystal (LiInS$_2$). In the process a few-cycle dispersive wave at $5.0\mic$ was generated with an efficiency of around 10\%. For a longer input pulse and a higher intensity we showed that a supercontinuum forms after 15 mm of propagation in this crystal. For input wavelengths around $3.0\mic$ the supercontinuum spans from $1.0-8.0\mic$, nearly 4 octaves, and the simulations showed that the blue part of the supercontinuum originates from strong spectral broadening of the SH. 
These examples highlight the enormous potential for using cascaded SHG in mid-IR nonlinear crystals for generating and shaping ultrafast and ultrabroadband mid-IR pulses.

Acknowledgments: the Danish Council for Independent Research, projects no. 274-08-0479 and no. 11-106702.

\end{document}